%% file: paper.tex
\documentclass[pra,graphicx,reprint,nobibnotes,amsmath,superscriptaddress,groupedaddress]{revtex4-1}
\usepackage[utf8]{inputenc}
\usepackage{mathtools}
\usepackage{amsmath}
\usepackage{graphicx}
\usepackage{hyperref}
\usepackage{siunitx}
\usepackage{textcomp}
\usepackage{physics}
\usepackage{float}
\DeclareMathOperator{\env}{env}

\begin{document}

\title{On the control of Rydberg state population with realistic femtosecond laser pulses}

\author{Janne Solanpää}
\email[]{janne@solanpaa.fi}

\author{Esa Räsänen}
\email[]{esa.rasanen@tut.fi}

\affiliation{Laboratory of Physics, Tampere University of Technology,
             Tampere FI-33101, Finland}

\date{\today}

\begin{abstract}
\noindent We investigate computationally a method for ultrafast preparation of alkali metal atoms in their Rydberg states
using a three-dimensional model potential in the single active electron approximation. By optimizing laser pulse shapes
that can be generated with modern waveform synthesizers, we propose pulses for controlling the population transfer from the ground state to a preselected set of Rydberg states.
Dynamical processes under the optimized pulses
are shown to be much more complicated than in the traditional optical two-photon preparation
of Rydberg states.
\end{abstract}

\maketitle

\section{Introduction}

\noindent Rydberg states have been observed in numerous systems
including, e.g., alkali metal atoms~\cite{[{See, e.g., }][{ and the references therein.}]alkali_rydberg_lifetimes_calculation}
and larger systems such as water~\cite{water_rydberg_experiment} and NO molecules~\cite{no_rydberg}.
Their features include long lifetimes~\cite{alkali_rydberg_lifetimes_calculation},
macroscopic extent of the electron wave function, and large
dipole moments~\cite{rydberg_atoms_book}. These features make them prime candidates for applications in, e.g.,
quantum information and quantum computing~\cite{rydberg_quantum_information}. They are also of
fundamental interest in the study of quantum chaos~\cite{luukko_rydberg}.

Experimental preparation of isolated alkali metal atoms in Rydberg states
can be achieved with a two-photon absorption~\cite{rydberg_atoms_book}. In rubidium,
the successive absorption of 480~\si{nm} and 780~\si{nm} photons can excite the valence electron
to a high-$n$ Rydberg state with up to 80\,\%  probability~\cite{saffman_experiment}.
However, the two-photon absorption technique requires (i) tuning of the laser frequencies to the desired
resonances and (ii) long irradiation durations to achieve reasonable yields~\cite{saffman_experiment}.

Addressing these drawbacks may be achieved by using laser pulses with tailored temporal profiles.
Standard techniques exist for the production of tailored femtosecond laser pulses~\cite{[{See, e.g., }][{ and references therein.}]hassan_rev_sci_instrum},
and their applicability has been demonstrated for controlling various dynamical phenomena in atoms
such as above-threshold ionization~\cite{PhysRevA.93.013413,solanpaa_pes}
and high-harmonic generation~\cite{solanpaa_hhg,Castro2015,1674-1056-25-9-094213,0253-6102-65-5-601,PhysRevA.97.053414,RevModPhys.80.117,PhysRevA.91.063408}.
Population and excitation control of atoms with femtosecond pulses has been studied to some extent
both experimentally~\cite{*[{See, e.g., }][{ and references therein.}]Hornung2000,PhysRevA.65.043406,PhysRevA.68.041402,PhysRevA.70.063407},
and computationally~\cite{PhysRevA.65.043406,PhysRevA.68.041402}.
However, control of the excitation to high-$n$ Rydberg states using multicolor fields from modern light-field synthesizers has yet to be demonstrated.

In this work we investigate the applicability of tailored femtosecond laser pulses
to ultrafast excitation of alkali atoms to their Rydberg states.
Using a computational optimization scheme similar to Ref.~\cite{solanpaa_pes},
we optimize a set of experimentally feasible pulse parameters
and find optimal laser pulses that can achieve up to 20\,\% population transfer to the targeted states. The pulse durations
are typically less than a few dozen femtoseconds -- demonstrating the possibility of ultrafast Rydberg state preparation.

This paper is organized as follows. In Sec.~\ref{sec:numerical_methods}
we introduce the numerical methods and the optimization scheme. In Sec.~\ref{sec:results}
we discuss the optimal pulse shapes for ultrafast Rydberg state preparation and investigate
the underlying dynamical processes. Finally, in Sec.~\ref{sec:summary} we summarize our findings.

\section{Numerical methods}
\label{sec:numerical_methods}

\noindent As a prototype atom for optimization simulations,
we use lithium within the single active electron (SAE) approximation
with the static potential $V_0(r)$ introduced in Ref.~\cite{schweizer_model_potentials}.
The optimization scheme is independent of the precise atomic model, and
the scheme is readily applicable to other models of alkali metal atoms.
The laser-electron interaction is included in the dipole approximation,
yielding the velocity gauge
Hamiltonian (in Hartree atomic units~\cite{atomic_units_orig})
\begin{equation}
    \label{eqn:hamiltonian_vg}
    \hat{H}(t) = \frac{\hat{\mathbf p}^2}{2} + V_0(\hat{ \mathbf r} )
                 + A_z(t) \hat{p}_z,
\end{equation}
where we have restricted ourselves to linearly polarized laser fields.

Our goal is to transfer the maximum amount of population from the initial state, 2s,
(with zero azimuthal quantum number, $m=0$)
to a certain set $\mathcal{I}$ of Rydberg states $\ket{\phi_{n,l}}$ (preserving $m=0$).
This can be achieved by maximizing the target functional
\begin{equation}
    \label{eqn:target_functional}
    G[\mathbf u] = \sum\limits_{\ket{\phi_{n,l}}\in\mathcal{I}}
        \left \vert \bra{\phi_{n,l}}\ket{\psi(T_{\max})} \right \vert^2,
\end{equation}
where $\mathbf u$ is the set of optimizable parameters,
and $\ket{\psi(T_{\max})}$ is the electron state at the end of the laser pulse.

The optimizable parameters $\mathbf u$ define the temporal shape of the laser
vector potential $A_z[\mathbf u](t)$.
Similarly to the approaches in Refs.~\cite{solanpaa_pes,hhg_bodi}, the
pulse is constructed as a superposition of multiple
\emph{channels}, each with a single central wavelength, i.e.,
\begin{equation}
A_z(t) = \sum\limits_{i=1}^N \frac{A_i}{\omega_i}
                            \env(t - \tau_i, \sigma_i)
                             \cos\left[ \omega_i(t-\tau_i)+\phi_i\right],
\end{equation}
\begin{figure}
    \includegraphics[width=\linewidth]{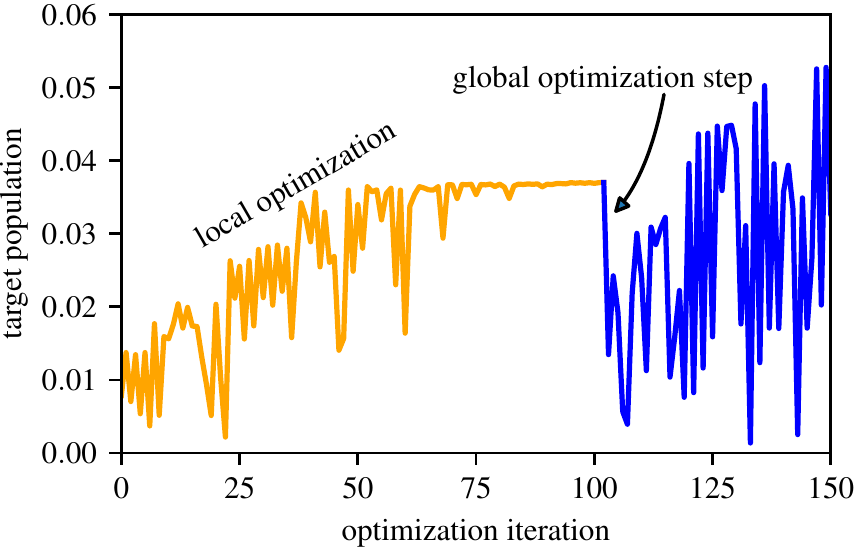}
    \caption{Total population of the target states as a function of the
    		optimization iterations demonstrating the working principle of the two-level
             optimization scheme. After convergence of the first local optimization (orange curve),
             the global optimizer restarts the local optimizer in a different region of the search space
             (at iteration number 102). Here we have targeted the states \mbox{$n=7$, $l=0\dots2$}.}
    \label{fig:optimization_process}
\end{figure}
where $A_i$, $\omega_i$, $\tau_i$, $\phi_i$, and $\sigma_i$ are the
amplitude, frequency, time of envelope maximum, carrier-envelope phase, and field FWHM
of each channel. The channel envelope is given b
\begin{equation}
\env(t - \tau, \sigma) = \left\{\begin{array}{cl}
        \exp\left[
            -\frac{\log(2)}{1-\left(\frac{t-\tau}{2\sigma}\right)^2}
                \left(\frac{t-\tau}{\sigma}\right)^2 \right],
                        & \vert t-\tau \vert < 2\sigma \\
            0, & \text{otherwise}
                    \end{array}\right.
\end{equation}
This is
a modified Gaussian which goes to zero at twice the FWHM,
and it is infinitely times differentiable
everywhere. This pulse parametrization allows us to model realistic pulse shapes
that can be generated with modern light field synthesizers~\cite{hassan_rev_sci_instrum}.

Calculation of the target functional in Eq.~\eqref{eqn:target_functional} for each pulse shape
requires us to (i) compute stationary states $\ket{\phi_{n,l}}$ of the system
and (i) propagate the initial state of the system under the laser vector potential.
The stationary states are obtained by
solving the effective radial equation for each angular quantum number $l$
with first-order finite differences. Time propagation of the electron wave function
is carried out with the \texttt{QPROP}
software, version 2.0~\cite{qprop} using
the Crank-Nicolson scheme~\cite{crank_nicolson}. For simulation parameters,
we have used the radial grid spacing \mbox{\num{0.1} a.u. (\num{0.005} \si{nm})}, radial grid length \mbox{\num{300} a.u. (\num{16} \si{nm})}, 
$l$ quantum numbers up to \num{50}, imaginary
absorbing potential of width \mbox{\num{50} a.u. (\num{2.6} \si{nm})}, and time-step
\mbox{\num{0.02} a.u. (0.5 \si{as})} for the simulations. Convergence of a few selected results was checked with higher accuracy.

The \texttt{QPROP} software was modified and wrapped for use within
\mbox{Python 3~\cite{python36}} for
interfacing with the optimization library \texttt{nlopt}~\cite{nlopt}.
Optimization is performed with a two-level scheme:
global optimization is carried out with \emph{multi-level single-linkage} (MLSL) algorithm~\cite{mlsl}
which essentially restarts local optimization while avoiding
previously found local extrema~\cite{nlopt}, and for
local optimization we use the derivative-free, trust region -based algorithm
called the \emph{Constrained Optimization BY Linear Approximations} (COBYLA) method~\cite{cobyla,nlopt}.
This derivative-free optimization scheme does not require the computation of the gradient
of Eq.~\eqref{eqn:target_functional}.

The optimization routine is provided with one to six different channels
with fixed central wavelengths
\mbox{300 \si{nm}}, \mbox{400 \si{nm}}, \mbox{800 \si{nm}}, \mbox{700.2243 \si{nm}},
\mbox{1.6 \si{\micro m}}, and
\mbox{2 \si{\micro m}}.
The value close to \mbox{700 \si{nm}} is in resonance with the 2s$\to$ 2p transition.
Furthermore, each channel is constrained to maximum electric field amplitude of
\mbox{$A_i\leq$ \num{67} \si{GV/m}}.
The time of the envelope maximum is allowed to vary \mbox{$\pm$ 6 \si{\femto s}},
and the field FWHM of each channel can have values
between \num{2.4} \si{\femto s} and \num{15} \si{\femto s}.

\section{Optimization results}
\label{sec:results}

\begin{table}
\caption{Summary of maximum achieved target populations for different pulse
channel combinations.}
\vspace*{\baselineskip}
\include{results_summary}
\label{tbl:summary}

\end{table}

\noindent A typical optimization process is shown in Fig.~\ref{fig:optimization_process}.
It begins with a random initial pulse within the constrained
search space, and local optimizer looks for a local maximum of the target functional.
After a local maximum has been found, approximately at iteration number 102, the global optimization routine
takes over and provides the local optimizer a new initial guess
causing a sharp drop in the target value. A typical optimization
simulation runs approximately 100 to 200 optimization steps providing up to a few local maxima. 
Optimization of the pulse parameters 
is indeed crucial for reaching reasonable target populations. In the example of Fig.~\ref{fig:optimization_process}
the optimization starts with a random pulse combination reaching barely 1\,\% target population, but the
optimization shapes the pulse to provide up to 5\,\% target population (at iteration number 149).

The best results for each set of target states are collected in Table~\ref{tbl:summary}.
We only show the best one or two channel combinations for each target,
but all possible channel combinations were tested. The optimized target populations range from 
90\,\% for the simplest target down to 3\,\% for more difficult to reach target states such as $n=7$, $l=4\dots6$.
We have also investigated the excitation of the system to a single target state: for 7f we have reached up to 2.5\,\% population
and for 8i up to 1.7\,\%. These moderate populations of single target states suggest the scheme lacks the finesse to target single Rydberg states.
However, due to finding only a few
local extrema per target for each channel combination, it may be possible to
improve these results with more optimization simulations and/or gradient based
algorithms. 

The maximum populations in Table~\ref{tbl:summary} are less than 
those achieved in previous works on optimal control 
of population transfer in atoms and molecules, e.g., in Refs.~\cite{rabitz_oct_of_population,SUGAWARA2002290}.
However, one must take into account the extremely constrained pulse combinations required by 
modern waveform synthesizers. In particular, the pulses with fixed channel wavelengths 
lack the ability to play with the resonances of the system.
Moreover, the Gaussian envelope of each channel forbids any sudden changes in the temporal profiles of the pulses, 
and shortness of the resulting pulses forces the control scheme to consider multiple complex transitions between the states.

\begin{figure}
    \includegraphics[width=\linewidth]{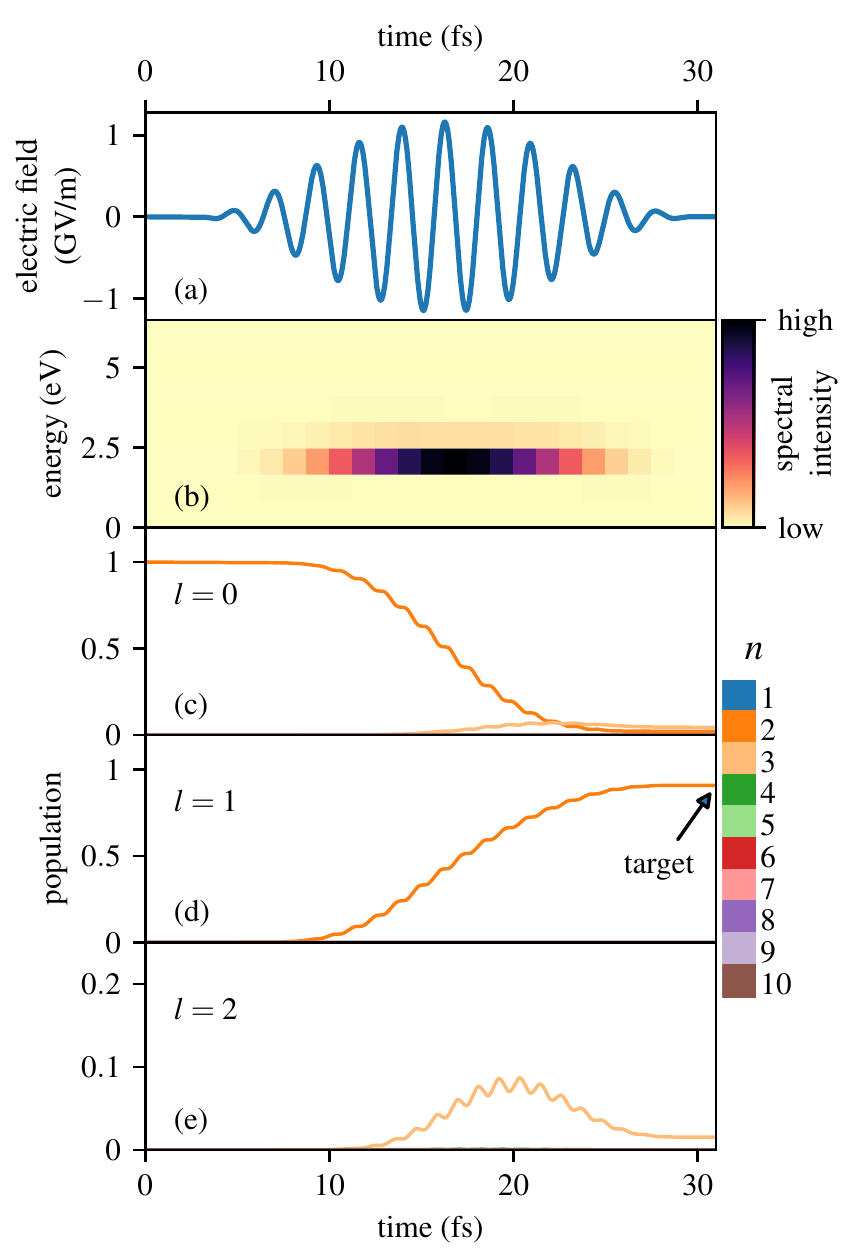}
    \caption{(a) The optimized laser pulse
    for populating the 2p state, (b) the power spectral density of the laser electric field, and (c)-(e)
    the populations of the stationary states.}
    \label{fig:target_2p}
\end{figure}

Next we will inspect the population transfer mechanisms behind the optimal pulses for a few select examples from Table~\ref{tbl:summary}.
The simplest transition to consider is \mbox{2s $\to$ 2p}. This is forbidden
for hydrogen, but for
Li the transition is allowed. This transition also serves as the first step in the optical preparation of Rydberg states
through a two-step excitation~\cite{rydberg_atoms_book}.
We find the optimal population transfer to be achieved with a pulse consisting only of the 700 \si{nm} channel
-- not surprising since the channel is in resonance with the transition. The pulse and the populations of the
few lowest states are show in Fig.~\ref{fig:target_2p}. The optimal pulse has a small peak electric field to avoid ionization,
and a 90 \% population transfer is  achieved with a pulse duration (intensity FWHM) less than $10$ fs. Merely increasing the pulse duration would not improve
the result since the initial state, 2s, is already depleted with the current pulse shape.
At first, the population transfer \mbox{2s $\to$ 2p} seems like a simple few-level
process. Indeed, a two-level model with the states 2s and 2p under the laser pulse of Fig.~\ref{fig:target_2p}
already yields an 80\,\% population transfer. However, even a bound-state model with all states up to $n=10$
fails to reach the 90\,\% yield of the full model. This suggests either the involvement of very high Rydberg states
or perhaps even the continuum in the full population transfer.

\begin{figure}
    \includegraphics[width=\linewidth]{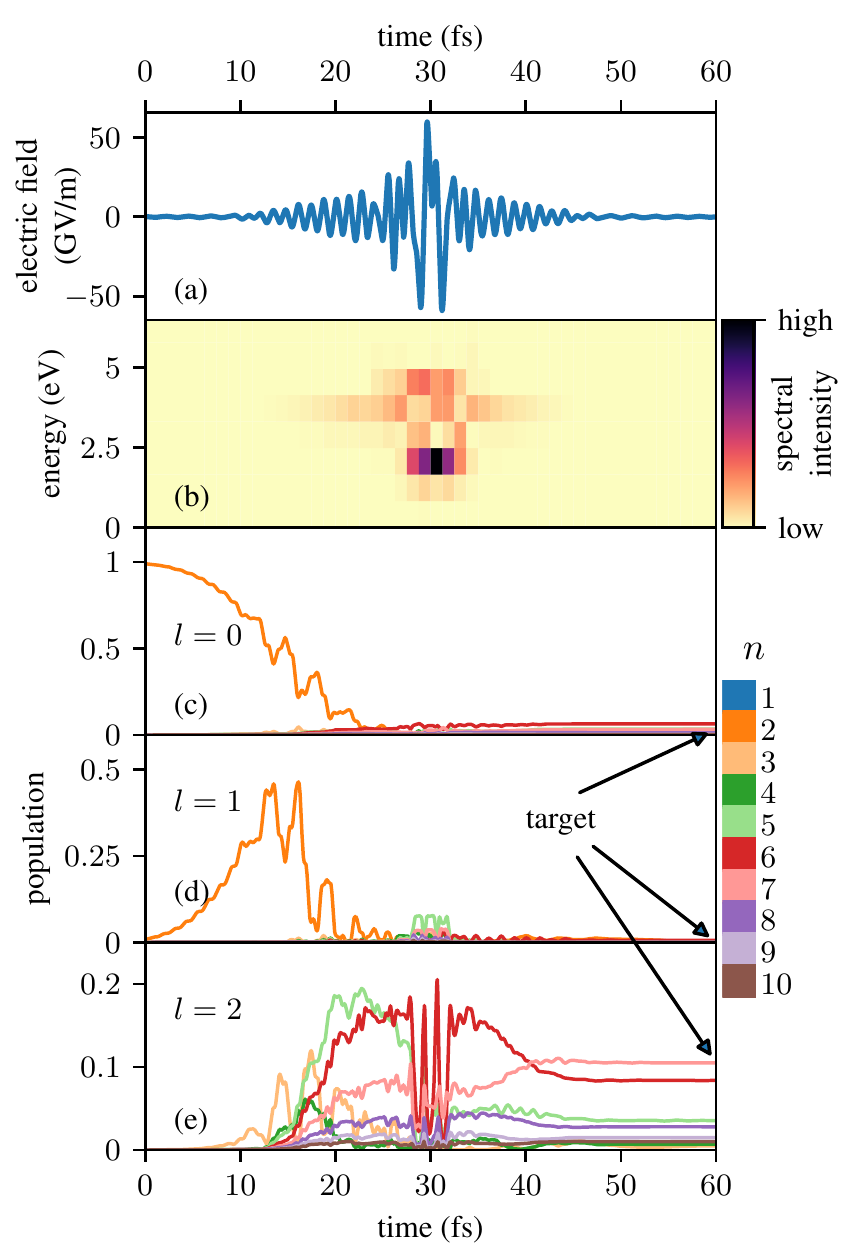}
    \caption{(a) The optimized laser pulse for populating the set of states $n=7$, $l=0\dots2$
    using the channels 
    \mbox{800 \si{nm}}, \mbox{700 \si{nm}}, \mbox{400 \si{nm}}, and \mbox{300 \si{nm}},
    (b) the power spectral density of the electric field,
    and (c)-(e) the populations of the stationary states.}
    \label{fig:target_7spd}
\end{figure}

Let us turn our attention to ultrafast population of Rydberg states. Targeting the states $n=7$, $l=0\dots2$,
our scheme yields a solid 14\,\% final population using the channels 800 \si{nm}, 700 \si{nm}, 400 \si{nm}, and 300 \si{nm} (see Fig.~\ref{fig:target_7spd}).
These channels are mixed with peak electric field ratios of $35 : 1 : 13 : 16$. While the 700 \si{nm} channel
is relatively weak compared to others, it is of utmost importance. and without it the final target population would drop to 0.2\,\%.
The optimized population transfer is somewhat akin to the traditional two-step excitation: first the electron is excited
from 2s to 2p by the weak 700 \si{nm} component; however, the second step is a much more complicated process involving multiple transitions
resulting in most of the final target population in the 7d state.

\begin{figure}
    \includegraphics[width=\linewidth]{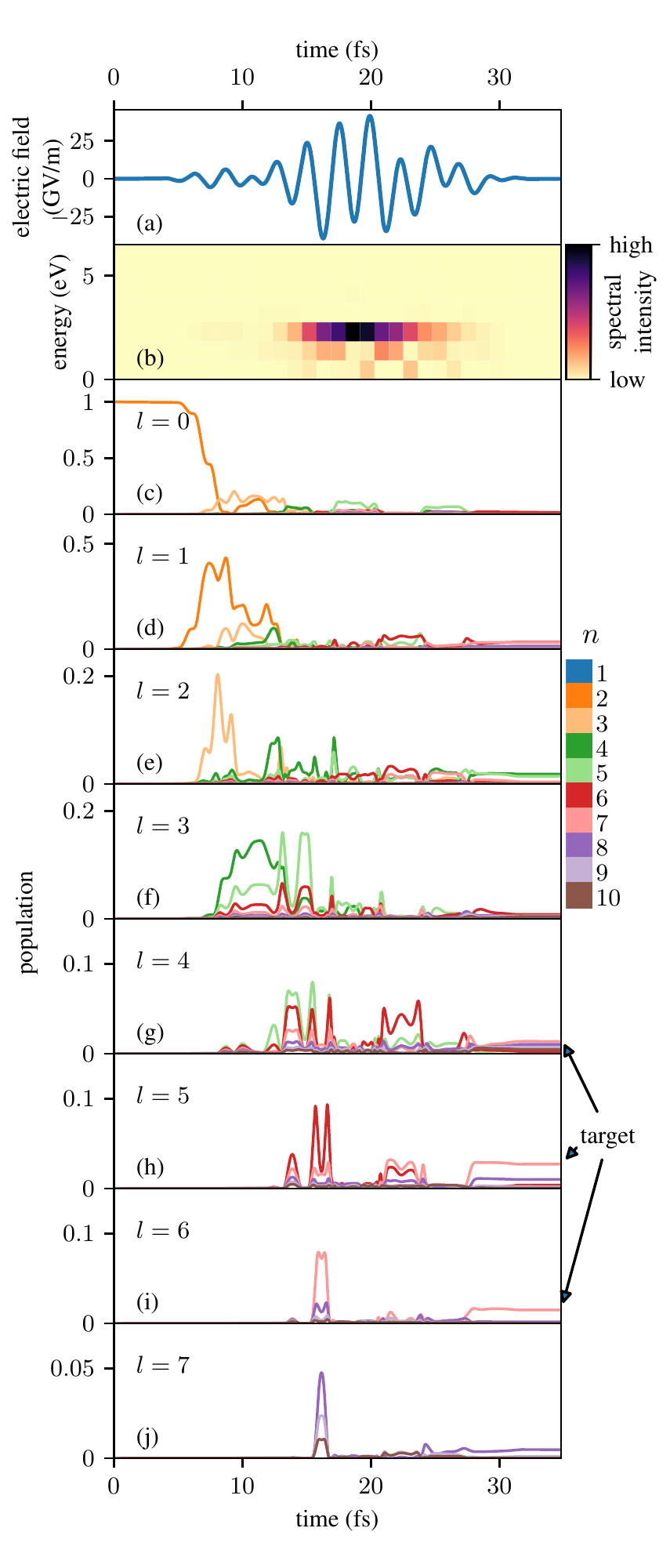}
    \caption{(a) The optimized laser pulse for populating the set of states $n=7$, $l=4\dots6$ using the channels 
    \mbox{2 \si{\micro m}}, \mbox{800 \si{nm}}, \mbox{700 \si{nm}}, and \mbox{400 \si{nm}},
    (b) the power spectral density of the electric field,
    and (c)-(j) the populations of the stationary states.}
    \label{fig:target_7ghi}
\end{figure}

Next, we will focus our attention to a more complicated target, $n=7$ $l=4\dots6$, which can not be reached with two-photon absorption, in contrast with the previous example.
An optimized pulse of duration less than 30 \si{fs} can transfer up to 6\,\% of the electron population to the target states.
The pulse, shown in Fig.~\ref{fig:target_7ghi}(a), mixes the channels
2 \si{\micro m}, 800 \si{nm}, 700 \si{nm}, and 400 \si{nm}
in ratios of electric field peak amplitude as $1:1.8:2.25:0.04$,
The 800 \si{nm} and 700 \si{nm} channels activate simultaneously, while the 2 \si{\micro m}
channel activates 6 \si{fs} later than the previous ones. The channels overlap significantly in time
yielding a complicated process for the population transfer. First few femtoseconds, up to approximately
$t=12$ \si{fs} transfer the population from 2s to highers states with $l\approx1\dots 3$
whereas the rest of the pulse makes the electron population oscillate between multiple states
and partly ionize.

\begin{figure}[t!]
    \includegraphics[width=\linewidth]{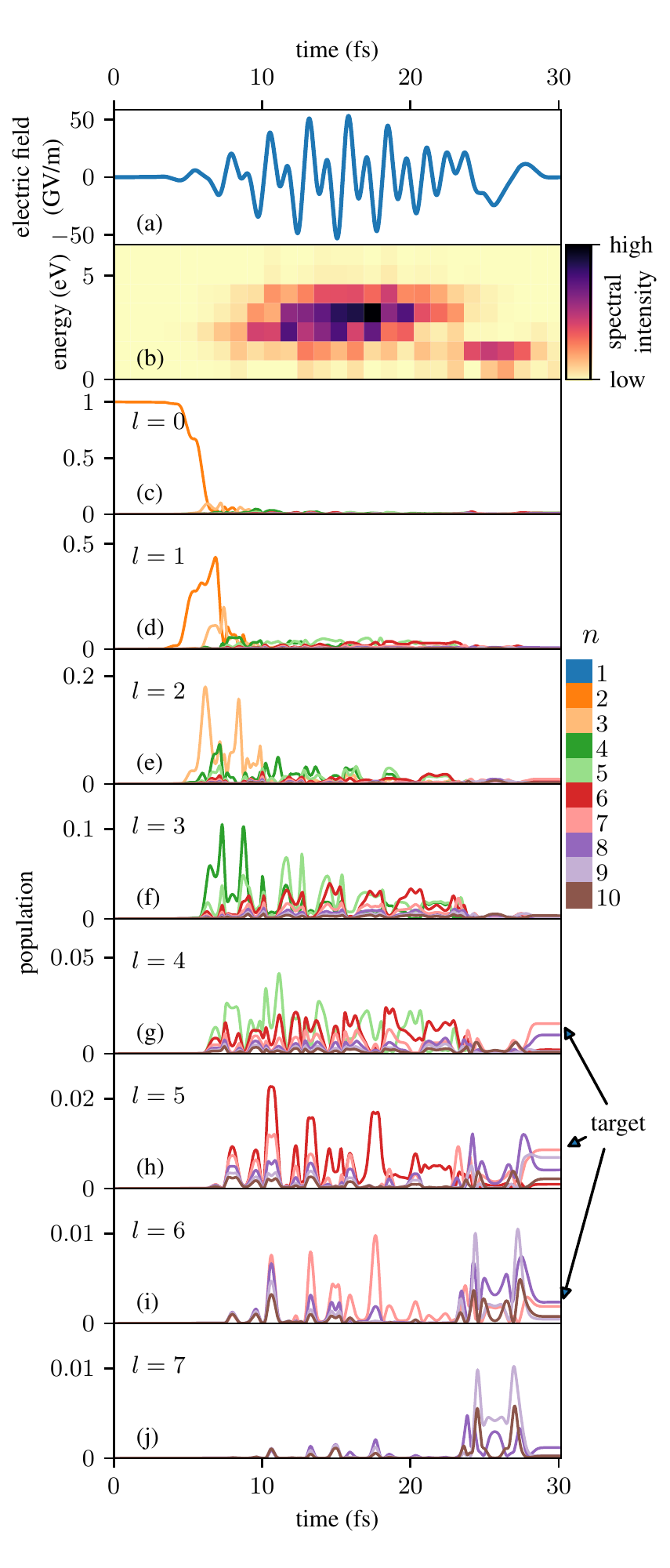}
    \caption{Sames as Fig.~\ref{fig:target_7ghi}, but without the \mbox{700 \si{nm}} channel.}
    \label{fig:target_7ghi_wo700}
\end{figure}

A question arises, whether the 700 \si{nm} pulse is an essential primer to achieve
the initial 2s $\to$ 2p excitation. This is not the case, as demonstrated in Fig.~\ref{fig:target_7ghi_wo700}
where we optimize the same target as previously but \emph{without} the 700 \si{nm} channel. A good initial population transfer to the 2p
state can still be found; however, rest of the population transfer process is naturally different due to different pulse temporal shape.

\begin{figure}
    \includegraphics[width=\linewidth]{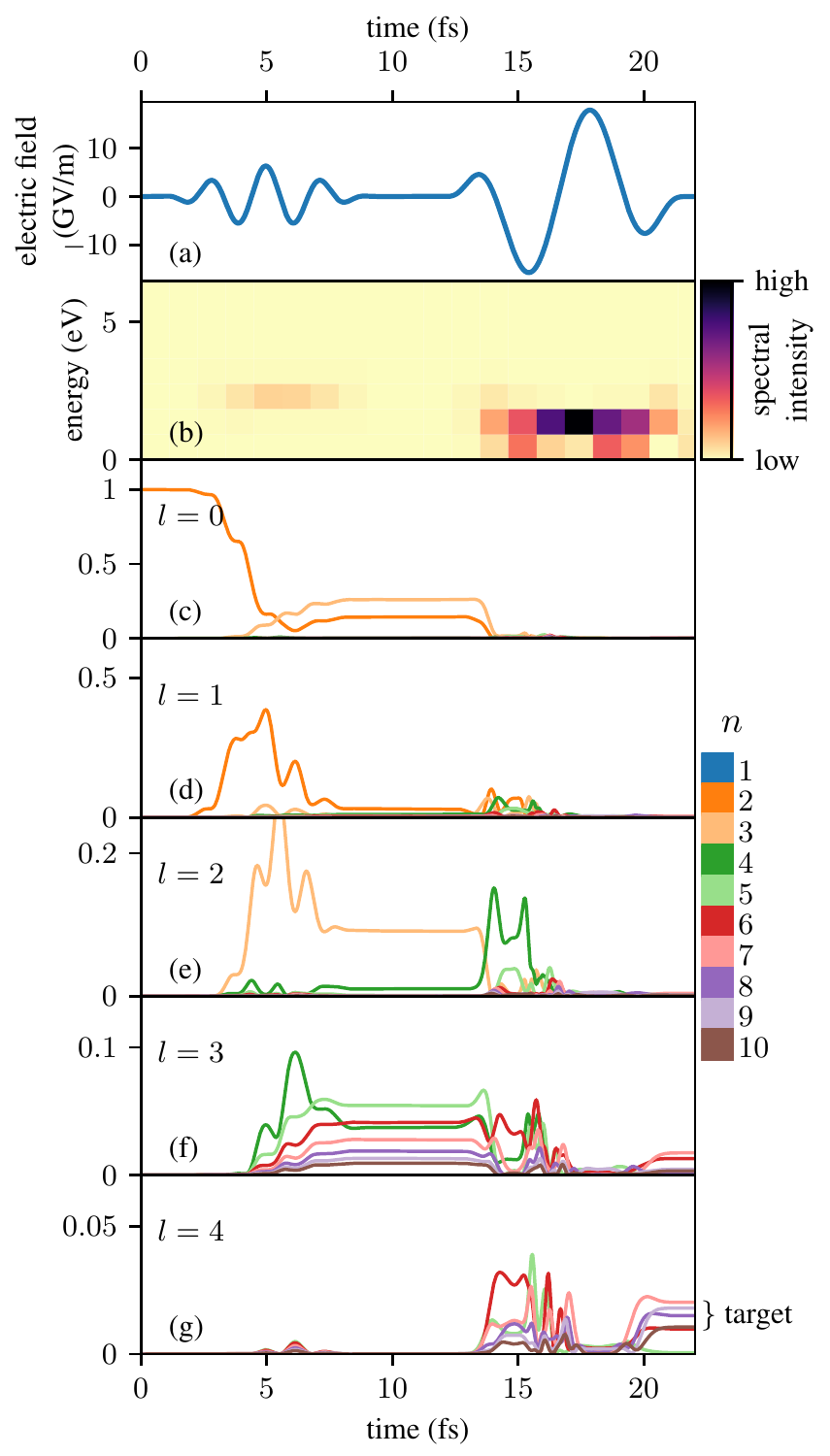}
    \caption{(a) The optimized laser pulse for populating the set of $l=4$ states with principal quantum numbers $n=7\dots10$ 
    consists of first a few-cycle \mbox{700 \si{nm}} primer followed by single cycle \mbox{2 \si{\micro m}} pulse,
    (b) the power spectral density of the electric field,
    and (c)-(g) the populations of the stationary states.}
    \label{fig:target_l4}
\end{figure}

We will now turn our attention to targeting the population of a single angular quantum number, e.g., $l=4$ with $n=7\dots10$.
The optimal pulse, shown in Fig.~\ref{fig:target_l4}(a) is a sequence of 700 \si{nm} and 2 \si{\micro m} channels providing us final
target population of 6\,\%. To analyze the population transfer process via pair-wise transfer rates,
notice first that the state populations $\vert c_{n,l} \vert^2 = \vert \bra{\phi_{n,l}}\ket{\psi(t)} \vert^2$ are equivalent
in the Schrödinger and interaction pictures of quantum mechanics.
In the interaction picture, the expansion coefficients obey the system of ordinary differential equations~\cite{[{See, e.g., }][{ or any other standard book on elementary quantum mechanics.}]sakurai}
\begin{equation}
\begin{split}
    \frac{\mathrm{d}}{\mathrm{d}t} c_{(n,l)}^I(t) =& -i A_z(t) \sum_{(n',l')} W_{(n,l),(n',l')} \\
    &\times\exp\{i [E_{(n',l')}-E_{(n,l)}] t\} c_{(n',l')}^I(t)m,
\end{split}
\end{equation}
where $W_{(n,l),(n',l')}$ is the z-component of the $(n,l),(n',l')$ momentum matrix element in the Schrödinger picture.
Now the pair-wise transfer rates are given by
\begin{equation}
    T_{(n,l),(n',l')}(t) = A_z(t)^2 \vert W_{(n,l),(n',l')} \vert \bra{\phi_{n,l}^S}\ket{\psi^S(t)} \vert^2.
\end{equation}

The transfer rates $T$ for the optimized population transfer to $l=4$, $n=7\dots10$
are shown as a function of time in the animation that can be found in Supplementary Material.
The first, 700 \si{nm} pulse excites the 

system from the initial 2s state to a set of $l=3$ states ($n=4\dots10$) via 2p.
Some population is left in the 2p and 3d states. The second, 2 \si{\micro m} pulse first transfers population leftovers from the 2p state
via 4d state to the $f$-states, and after its first optical cycle, the second pulse transfers the population from the f-states to the targeted g-states.
Due to weak pulses, the system is essentially not ionized, but rest of the population escapes to higher bound states.
Transfer rates seem the obvious choice for interpreting the optimal population transfer processes for each target,
but they turn out to be significantly more complicated for most of the other targets.

\begin{figure}
    \includegraphics[width=\linewidth]{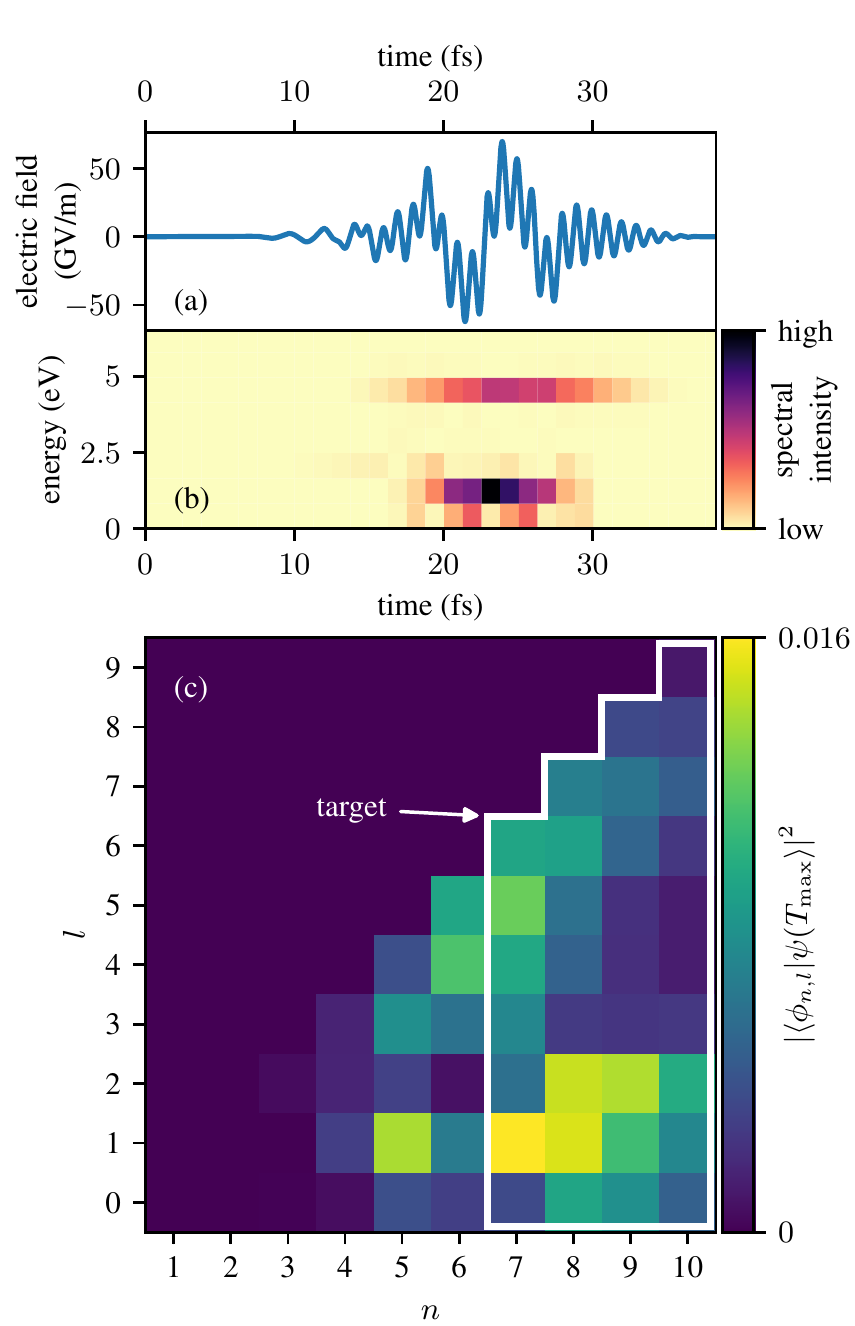}
    \caption{(a) The optimized laser pulse for populating the set of states $n=7\dots10$
    using the channels 2 \si{\micro m}, 800 \si{nm}, 700 \si{nm}, and 300 \si{nm},
    (b) the power spectral density of the electric field,
    and (c) the final populations of the stationary states.}
    \label{fig:target_big}
\end{figure}

As a final demonstration, we target the states with principal quantum numbers $n=7\dots10$ without restrictions to the angular quantum number.
Due to larger number of targeted states, the total target population reaches over 20\,\% with the optimal pulses with
the highest yield achieved with the channels 2 \si{\micro m}, 800 \si{nm}, 700 \si{nm}, and 400 \si{nm} shown in Fig.~\ref{fig:target_big}.
Most of the final target population is in low-$l$ states, peaking at 7p and 8p followed by their neighbours by coupling, 8d and 9d.\\

\section{Summary}

\label{sec:summary}

We have demonstrated the applicability of a few-color femtosecond pulses
realizable by modern waveform synthesis~\cite{hassan_rev_sci_instrum}
to optimal control of population transfer from ground state to a set of Rydberg states.
Our control scheme was found to achieve up to 23\,\% Rydberg-state populations
when transferring population to a few selected states, but when targeting a single state the
experimentally restricted pulse combinations do not seem to allow sufficient control over the excitation process.
Typical simulations with such realistic multicolor waveforms yield a complicated dynamical process
which usually can not be easily interpreted with clear excitation paths. In this respect, 
our results demonstrate a very different optimized dynamical process compared to having longer and less constrained pulses, which allows the exploitation of the resonances.

Further investigation would be warranted to study the applicability of our
scheme to, e.g., the preparation of circular Rydberg states, including field polarization as an additional control knob.
In addition, adoption of gradient-based optimization algorithms should
be straightforward and quite important especially when extending the demonstrated optimization
scheme to full multi-electron models with possibly even more complicated optimization landscapes.

\begin{acknowledgments}
\noindent The authors are grateful to M. Sarvilahti
for the idea regarding the modified Gaussian function.
This work was supported by the Academy of Finland (grants no. 267686 and 304458).
We also acknowledge CSC -- the Finnish IT Center for Science --
for computational resources.
\end{acknowledgments}

\bibliography{refs}

\end{document}

%% file: results_summary.tex
\setlength{\tabcolsep}{4pt}

\begin{tabular}{lll}
    Target               &   Channels (\si{\micro m})   &    Max. population    \\
    \hline\\
    2p                   &   0.7                                      &     91\,\%   \\
    \\
    $n=7$, $l=0\dots2$   &   0.8, 0.7, 0.4, 0.3      & 14\,\%        \\
                         &   0.8, 0.7, 0.4                   & 5\,\%            \\
    \\
    $n=7$, $l=4\dots6$   &   2, 0.8, 0.7, 0.4          &   6\,\%       \\
                         &   2, 0.8, 0.4                              &    3\,\%     \\
    \\
    $n=7\dots10$, $l=4$   & 2, 0.7                        &  6\,\%            \\
    \\
    $n=7\dots10$           & 2, 0.8, 0.7, 0.3             &  23\,\%             \\ 
                         &  2, 1.6, 0.8, 0.7, 0.3                &           21\,\%      
\end{tabular}